%% file: main.tex
\documentclass[acmtog]{acmart}
\makeatletter
\renewcommand\@formatdoi[1]{\ignorespaces}
\makeatother
\renewcommand\footnotetextcopyrightpermission[1]{} 
\usepackage{booktabs} 
\makeatletter
\def\runningfoot{\def\@runningfoot{}}
\def\firstfoot{\def\@firstfoot{}}
\makeatother

\setcitestyle{square}

\input{NewCommands}


\usepackage[T1]{fontenc}

\usepackage{amsmath,amsfonts,amssymb,amsthm,cancel,icomma,nicefrac,mathrsfs,
            eurosym,verbatim,environ,ifthen,ifdraft,pdfpages,float,booktabs}
\allowdisplaybreaks[1]
\usepackage[scaled=0.9]{DejaVuSansMono}

\usepackage{color}
\definecolor{lstgrey}{rgb}{0.95,0.95,0.95}
\usepackage{listings}
\begin{document}
\title{Taichi: An Open-Source Computer Graphics Library} 


\author{Yuanming Hu}
\affiliation{%
  \department{}
  \institution{MIT CSAIL}}
\email{yuanmhu@gmail.com}

\newcommand{\taichi}{{\em Taichi}}

\begin{abstract}
An ideal software system in computer graphics should be a combination of innovative ideas, solid software engineering and rapid development.
However, in reality these requirements are seldom met simultaneously.
In this paper, we present early results on an open-source library named \taichi{} (\url{http://taichi.graphics}) which alleviates this practical issue by providing an {\em accessible, portable, extensible, and high-performance} infrastructure that is {\em reusable} and {\em tailored} for computer graphics. As a case study, we share our experience in building a novel physical simulation system using \taichi.

\end{abstract}


\begin{teaserfigure}
\centering
\includegraphics[width=1.0\textwidth]{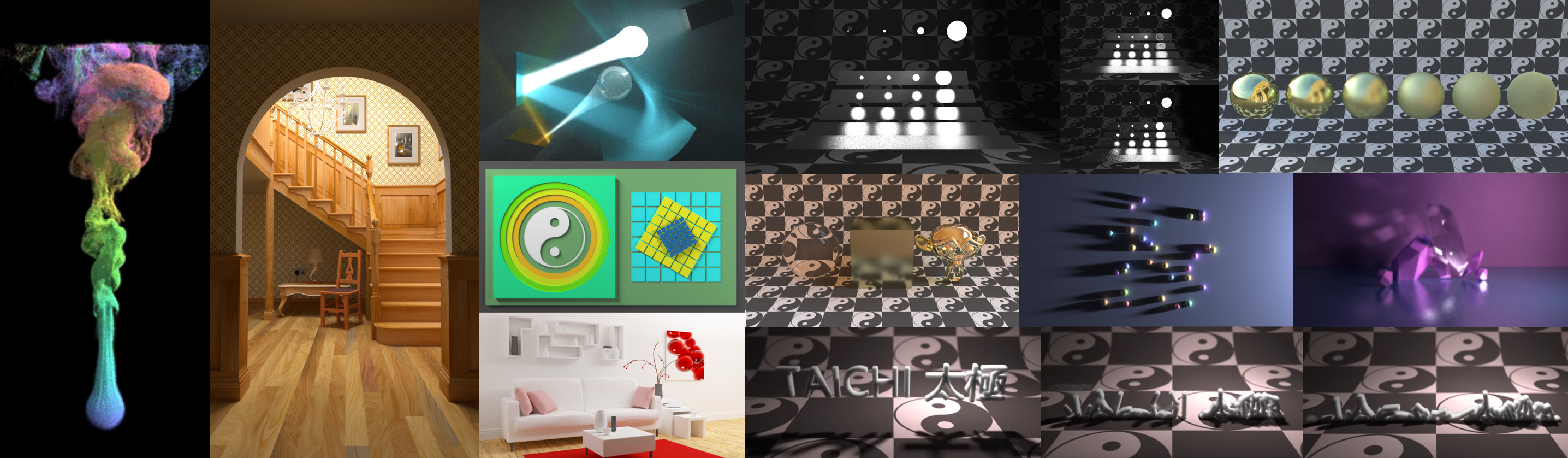}
\caption{Results simulated and rendered using \taichi.}
\label{fig:bridge}
\end{teaserfigure}

\maketitle

\section{Why New Library for Computer Graphics?}
Computer graphics research has high standards on novelty and quality, which lead to a trade-off between rapid prototyping and good software engineering.
The former allows researchers to code quickly but inevitably leads to defects such as unsatisfactory performance, poor portability and maintainability. 
As a result, many projects are neither reused nor open-sourced, stopping other people from experimenting, reproducing, and developing based on prior art. 
The latter results in reusable codes for future projects, but slows down research progress due to low-level engineering such as infrastructure building.
Admittedly, for a researcher too much low-level engineering may become an obstacle for high-level thinking.

Existing open-source projects are focused on certain functionality, like rendering (e.g. {\em Mitsuba}~\cite{mitsuba}, {\em PBRT} \cite{pharr2016physically}, {\em Lightmetrica}~\cite{lm15}, {\em POV-Ray}~\cite{povray}, etc.), geometry processing ({\em libIGL}~\cite{jacobson2013libigl}, {\em MeshLab}~\cite{cignoni2008meshlab}, {\em CGAL}~\cite{fabri2009cgal}, etc.), simulation ({\em Bullet}~\cite{coumans2013bullet}, {\em ODE}~\cite{smith2005open}, {\em ArcSim}~\cite{arcsim}, {\em VegaFEM}~\cite{sin2013vega}, {\em MantaFlow} \cite{mantaflow}, {\em Box2D}~\cite{box2d}, {\em PhysBAM}~\cite{dubey2011physbam}, etc.).

\begin{figure}[bh!]
\hspace{-0.7cm}
\includegraphics[width=1.03\linewidth]{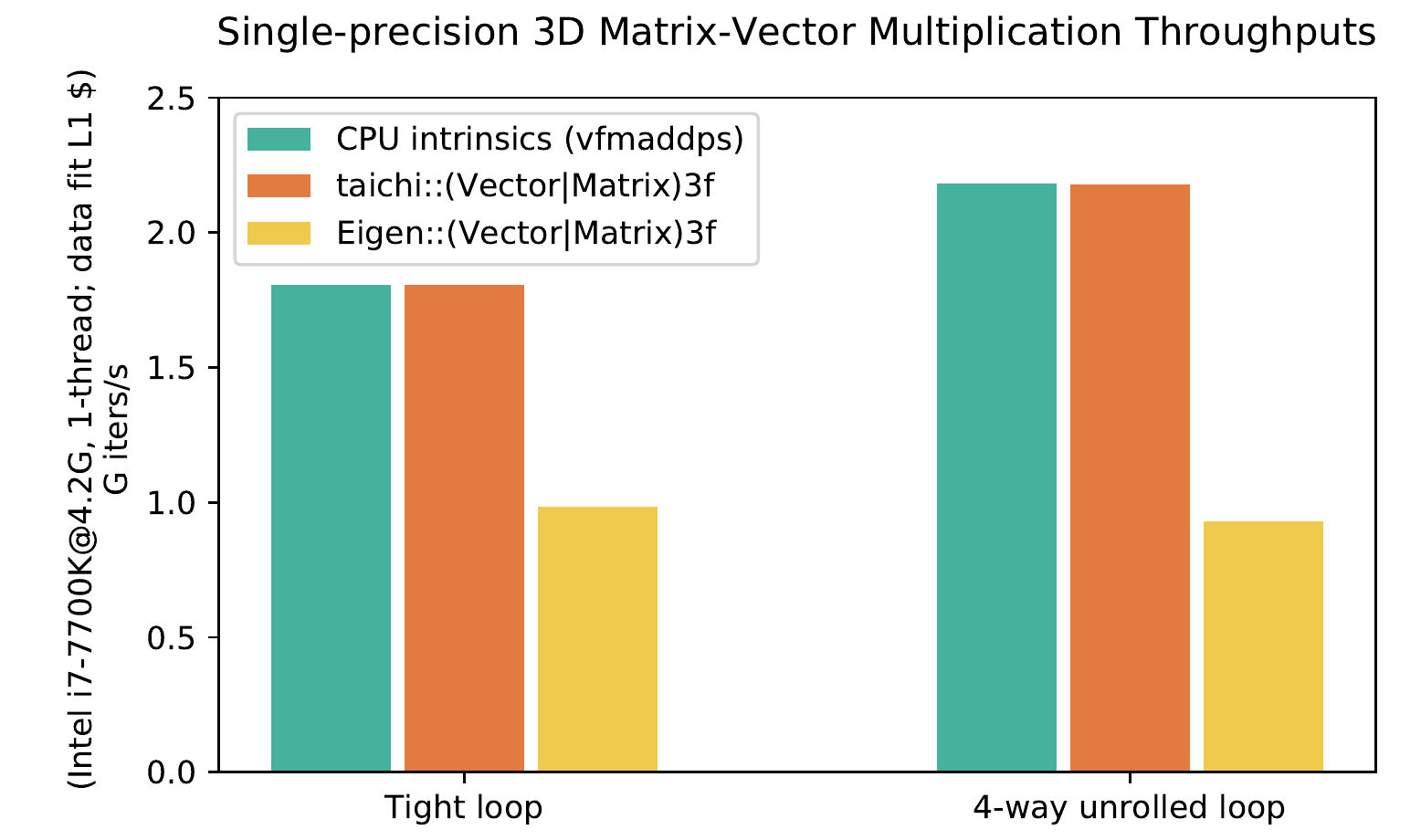}
\caption{\textbf{Why do we need a tool \underline{tailored} for graphics?} Consider $3$D matrix-vector multiplication as an example.
In Eigen single-precision 3D vectors are stored compactly using $12$ bytes.
Compared with $16$-byte-aligned storage, this scheme saves some space, but offers unsatisfactory performance since additional permutation in SIMD registers is needed for vectorized computation.
In graphics we care more about throughput and latency instead of space, so Taichi's vector system is designed in the latter way and is $1.8-2.3\times$ faster. }
\label{fig:eigen}
\end{figure}

The aforementioned libraries have proven successful in their own applications.
However, though some projects can more or less reuse these frameworks, building prototypical systems from scratch is often necessary when modifying these libraries becomes even more expensive than starting over.
\taichi{} is designed to be a {\em reusable infrastructure} for the latter situation, by providing abstractions at different levels from vector/matrix arithmetics to scene configuration and scripting.
Compared with existing reusable components such as {\em Boost} or {\em Eigen}, \taichi{} is tailored for computer graphics. The domain-specific design has many benefits over those general frameworks, as illustrated in Fig.~\ref{fig:eigen} with a concrete example. 

\section{Design Goals}
\paragraph{Accessibility.}
\taichi{} is open-source and easy-to-use by design.
We avoid unnecessary dependencies, especially hard-to-deploy ones such as {\em Boost}. Installation can be done simply by executing a python script that automatically installs all dependencies.
For beginners, it contains many working demos that showcase how this software should be used at a high level.

\paragraph{Portability.}
\taichi{} is cross-platform and supports Linux, OS X and Windows.
Our aim is to make every project built on \taichi{} automatically portable by providing an abstraction of the actual operating systems.
This makes open-sourcing projects much easier.

\paragraph{Reusability.}
\taichi{} contains common utilities for developing graphics projects. Reusing them can significantly speed up development since the time spent on (re)inventing the wheels is saved. 

\paragraph{Extensibility.}
With a plugin-based design, \taichi{} allows users to quickly add their own implementations of existing or new interfaces. The implementations (``units") will be automatically registered at program load time and can be instantiated conveniently with a name identifier through the factory.

\paragraph{High-performance.}
We offer not only efficient (sometimes zero-cost) abstractions at different levels of \taichi, but also performance monitoring tools such as scoped timers and profilers to make performance engineering convenient.

\begin{figure*}[th!]
\centering
\includegraphics[width=1.0\textwidth]{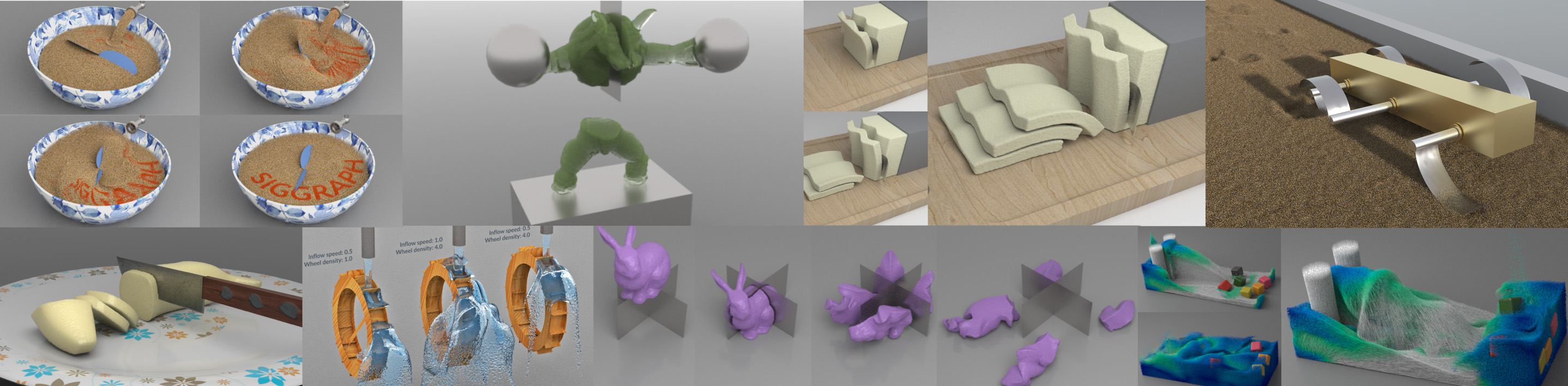}
\caption{Results in the MLSMPM-CPIC paper. Using Taichi as the code base significantly sped up the development of this project.}
\label{fig:mlsmpm}
\end{figure*}

\section{Components}

Here we present some representative components that can be reused for developing new computer graphics systems.

\paragraph{(De)serialization}
The serialization system allows marshaling internal data structures in \taichi{} into bit streams. A typical application is snapshotting. Many computer graphics programs take a long time to execute and it is necessary to take snapshots of current states. In case of run-time error, the program can continue running from the latest snapshot after the bug causing crashing is repaired. Other use cases include inter-program communication when there is no shared memory.
Our serialization library is header-only and can be easily integrated into other projects for reading the serialized \taichi{} data.

\paragraph{{Logging and Formatting}}
Appropriate logging is an effective way to diagnose a long-running program. \taichi{} internally uses {\em spdlog} and {\em fmtlib}. The former manages logging and the latter is a modern formatting library which unifies string formatting in C++ and Python. It is safer, easier to use and more portable than its alternatives like \lstinline|std::cout| and \lstinline|printf|.


\paragraph{{Profiling}}
Though there are mature external profiling tools like {\em gprof} or {\em Intel VTune}, using them needs additional manual operations. \taichi{} has an integrated scoped profiling system that records the time consumption of each part automatically, simplifying hotspot analysis.


\paragraph{{Debugging and Testing}}
C++ programs do not generate much useful debugging information when it crashes. This makes debugging hard and is the reason why \taichi{} captures run-time errors and prints stack back-trace. On Linux, it additionally triggers gdb for debugging. Once a task finishes (or crashes), \taichi{} can send emails to notify the user. \taichi{} uses {\em Catch2} 
for testing and we aim to build a high-coverage test suite for the library. 

\paragraph{Rendering}
Though not designed to compete with existing rendering frameworks, \taichi{} provides demonstrative implementations of popular rendering algorithms, from basic path tracing to modern ones such as vertex connection and merging~\cite{georgiev2012light}/unified path sampling~\cite{hachisuka2012path} with adaptive Markov chain Monte Carlo~\cite{vsik2016robust}. 

\paragraph{Simulation}
\taichi{} is shipped with several physical simulators including APIC~\cite{jiang2015affine}/FLIP~\cite{zhu2005animating} liquid/smoke simulation with a multigrid-preconditioned conjugate gradient (MGPCG)~\cite{mcadams2010parallel} Poisson solver for projection. We will also release a high-performance material point method code, as detailed in section~\ref{sec:case}.


\paragraph{{File IO support}}
\taichi{} supports reading and writing of popular file formats for computer graphics, including obj (via {\em tinyobjloader}) and ply, jpg, png, bmp, ttf (via {\em stb\_image, stb\_image\_write, stb\_truetype}
),  
In addition, \taichi{} wraps common utilities provided by {\em ffmpeg}. It can generate mp4 or gif videos from an array of images, either in memory or on disk.

\paragraph{Scripting}
\taichi{} has a hybrid design with a kernel part in \lstinline|C++14| and an easy-to-use interface in \lstinline|Python 3|. Such approach decouples inputs like demo set-up from the actual kernel computation code. 
{\em pybind11} is used for binding \lstinline|C++| interface for \lstinline|Python|.

In addition, we built \taichi{} based on {\em Intel Thread Building Blocks} for multithreading. An efficiently vectorized linear algebra library is also developed to suit computer graphics applications better, as mentioned in Figure~\ref{fig:eigen}. The \taichi{} developer team is in charge of maintaining and simplifying the deployment process of all the aforementioned dependencies of \taichi{}.



\section{Case Study: A Physical Simulation Project}
\label{sec:case}
So far, we have used \taichi{}  in five research projects. Published ones include ~\cite{mlsmpm, hu2017asynchronous}. Here we summarize our experience during the development of the SIGGRAPH 2018 paper ``A Moving Least Squares Material Point Method with Displacement Discontinuity and Two-Way Rigid Body Coupling" ~\cite{mlsmpm}, abbreviated as ``MLSMPM-CPIC". \taichi{} is used as the backbone of the simulator development of this project. Visual results are displayed in Figure~\ref{fig:mlsmpm}.

\paragraph{Productivity.}
All features mentioned in the previous section proved useful. In fact, many of them are developed during research projects like this one for productivity and future reusability. For example, the serialization system allows us to periodically save simulation states to disk, and conveniently restart from these snapshots. It is especially important for long-running simulations which can take hours or even days. Performance engineering is guided by the \taichi{} profiler, which shows a breakdown of time consumption and has led to significantly higher efficiency compared with the previous state-of-the-art \cite{tampubolon2017multi}.

\paragraph{Team Scalability.}
The python scripting system makes our development especially suitable for team working, because the algorithm development (in C++) and experimental validation (via python scripts) are nicely decoupled: 
for most of the time only one or two core members need to develop the C++ simulation code, and all team members can help conducting experiments without getting involved into low-level details.

In general, the team is satisfied with this infrastructure and has decided to build future projects on it. We believe the user experience will continue to improve as we battle-test it in more projects. 

\section{Future Work}


Clearly, a lot more engineering efforts are needed to improve \taichi. Detailed documentation and better test coverage are two urgent tasks. Improving and stabilizing the interface design, support for other open-source softwares like {\em Blender} are also very meaningful.

\section*{Acknowledgements}
The author would like to thank Toshiya Hachisuka and Seiichi Koshizuka for hosting his internship at the University of Tokyo, where he developed the initial version of \taichi. Chenfanfu Jiang provided helpful suggestions for developing \taichi{} and supported its adoption in several projects.
Other developers, especially Yu Fang, also contributed to \taichi. Some demo scenes are from~\cite{resources16}. Finally, thank all the open-source software developers for making their achievements freely available to everyone.

\bibliographystyle{ACM-Reference-Format}

\bibliography{main}

\section*{Change log}
April 24, 2018: initial version.

\end{document}

%% file: NewCommands.tex
\usepackage{mathrsfs}

\usepackage{multirow}
\usepackage{algorithm,algorithmic}
\usepackage{multirow}

\makeatletter
\def\hlinewd#1{%
\noalign{\ifnum0=`}\fi\hrule \@height #1 %
\futurelet\reserved@a\@xhline} 
\makeatother

\usepackage{graphicx}
\usepackage{amsmath}
\usepackage{amssymb}

\usepackage{parskip}
\usepackage{epstopdf}
\usepackage{tabularx}

\AppendGraphicsExtensions{.tiff}

\usepackage[labelfont=bf,textfont=it]{caption}
\usepackage{subfig}

\usepackage{algorithm} 
\usepackage{algorithmic}  

